\pdfoutput=1
\documentclass[fleqn,usenatbib]{mnras}

\usepackage{newtxtext,newtxmath}
\usepackage[T1]{fontenc}

\DeclareRobustCommand{\VAN}[3]{#2}
\let\VANthebibliography\thebibliography
\def\thebibliography{\DeclareRobustCommand{\VAN}[3]{##3}\VANthebibliography}

\usepackage{graphicx}	% Including figure files
\usepackage{amsmath}	% Advanced maths commands
\title[PBH dark matter strong lensing constraints]{Strong lensing constraints on primordial black holes as a dark matter candidate}

\author[V. Dike et al.]{
Veronica Dike,$^{1}$\thanks{E-mail: veronica.j.dike@gmail.com}
Daniel Gilman,$^{2}$
and Tommaso Treu$^{1}$
\\
$^{1}$Department of Physics and Astronomy, University of California Los Angeles, Los Angeles, CA, 90095, USA\\
$^{2}$Department of Astronomy and Astrophysics, University of Toronto, 50 St. George Street, Toronto, ON, M5S 3H4, Canada
}

\date{Accepted XXX. Received YYY; in original form ZZZ}

\pubyear{2022}

\begin{document}
\label{firstpage}
\pagerange{\pageref{firstpage}--\pageref{lastpage}}
\maketitle

\begin{abstract}
Dark matter could comprise, at least in part, primordial black holes (PBH). To test this hypothesis, we present an approach to constrain the PBH mass ($M_{\rm{PBH}}$) and mass fraction ($f_{\rm{PBH}}$) from the flux ratios of quadruply imaged quasars. Our approach uses an approximate Bayesian computation (ABC) forward modeling technique to directly sample the posterior distribution of $M_{\rm{PBH}}$ and $f_{\rm{PBH}}$, while marginalizing over the subhalo mass function amplitude, spatial distribution, and the size of the lensed source. We apply our method to 11 quadruply-imaged quasars and derive a new constraint on the intermediate-mass area of PBH parameter space $10^4 $M$_{\odot}<M_{\rm{PBH}}<10^6$M$_\odot$. We obtain an upper limit $f_{\mathrm{PBH}}<0.17$ (95\% C.L.). This constraint is independent of all other previously published limits.
\end{abstract}


\section{Introduction}

Primordial black holes (PBH) are an appealing DM candidate because they do not require physics beyond the standard model and black holes are known to exist in nature. In the early Universe, overdensities could have created the earliest black holes (\citealt{Zeldovich1967}; \citealt{Hawking1971}), and these black holes could persist into the present day to make up part or all of the dark matter (DM) in the Universe (\citealt{CarrHawking1974}; \citealt{Chapline1975}). For a recent overview of  primordial black holes as a DM candidate, see \cite{CarrDM2020} and \cite{Green2021}. Narrowing the primordial black hole  parameter space can place constraints on various models of cosmological significance because PBH formation and evolution is entwined with the history of the Universe. 
	
The hypothetical parameter space for PBH is very wide. Their mass distribution is virtually unconstrained theoretically, and they do not necessarily have to account for the entirety of DM. Observations exclude many mass ranges for PBH representing 100\% of DM, but the constraints are much weaker for a population that makes up only a fraction of DM (see, e.g., \citealt{Belotsky2019}; \citealt{CarrConstraint2020}). For black holes of mass $M_{\mathrm{PBH}}$ greater than $10^2$M$_\odot$, fast radio burst lensing constrains the fraction of dark matter in PBH, $f_{\mathrm{PBH}}$, to less than 9\% (\citealt{Zhou2021}). PBH as all dark matter requires a population of around sub-solar mass, but is consistent with the FRB rate in the model of \cite{Kainulainen2021}. A stellar-mass scale PBH distribution is detectable as microlensing in strongly lensed quasars, in cases where microlensing by stars can be suppressed \citep{Hawkins2020}.

\cite{CarrConundra2019} put forward a multi-peaked PBH mass function (including a high-mass peak of 10$^6$ M$_\odot$) that could explain a range of phenomena from cosmic infrared background excess to black hole merger rates. See also \cite{Khlopov2010} for an overview of cosmological implications of PBH formation mechanisms. PBH non-detection can itself constrain the scale of isocurvature perturbations in cold dark matter in the early Universe \citep{Passaglia2021}. A small DM fraction of high-mass PBH could seed supermassive black holes and galaxy formation \citep{CarrStructure2018}, and in turn the observed population of supermassive black holes can also be used to constrain the PBH mass function \citep{Cai2023}.

A powerful and direct way to probe the PBH contribution to dark matter is strong gravitational lensing (see, e.g., \citealt{Treu2010} and references therein). Intrinsically point masses, PBH are particularly effective deflectors. Their observational signature depends only on their mass, with the deflection angle in terms of impact parameter $\xi$ modeled as ${\alpha} = 4GMc^{-2}\xi^{-1}$.  The method of gravitational imaging \citep{Koopmans2005,Vegetti2010,He2022}could in principle be used to detect individual PBH of masses greater than 10$^3$ M$_\odot$ \citep{Banik2019}, and lensing constraints from compact radio sources can also be used to constrain high-mass PBH \cite{Zhou2022}. Because constraints from lensing are completely independent of others that have been used to constrain PBH in a similar mass range, such as dynamical constraints (\citealt{CarrDF1999}, \citealt{Quinn2009}, \citealt{Brandt2016}), X-ray background constraints on accretion rate \citep{InoueKusenko2017}, or Lyman-$\alpha$ forest enhancement constraints \citep{Afshordi2003, Mack2007, Murgia2019}, lensing provides a vital cross-check. 

Anomalies in the ratios of flux between images of the same lensed source can reveal substructure in the lensing mass distribution. This technique, suggested initially by \citet{MaoSchneider1998}, can probe structure at lower mass scales than those accessible with gravitational imaging. Such flux ratio anomaly studies rely on observations of lensed sources that are large enough to avoid being affected by stellar microlensing; see \cite{DoblerKeeton2006} for an analysis of the effect of source size on flux ratio analysis. Examples of such sources include radio emission in radio-loud quasars \citep{MaoSchneider1998, Metcalf2001, DalalKochanek2002, Hsueh2020}, mid-infrared emission from the hot dust in active galactic nuclei (AGN) \citep{Chiba2005}, and the the narrow-line region of AGN \citep{MoustakasMetcalf2003, Nierenberg2014, Nierenberg2017, Nierenberg2020}. 

\cite{Gilman2020, Gilman2020b} presented an analysis framework that uses the flux ratios among images in quadruply-imaged quasars (quads) to constrain the properties of dark matter structure in strong lens systems. These techniques can be adapted to constrain a variety of dark matter models, including cold dark matter (CDM), warm dark matter \citep{Gilman2020}, self-interacting dark matter \citep{Gilman2022b}, and fuzzy dark matter \citep{Laroche2022}, given a prescription for the halo mass function and density profiles of haloes.

In this paper, we present new constraints on the PBH parameter space by analysing the flux ratio anomalies in a sample of strongly-lensed quasars observed in the narrow-line regime \citep{Nierenberg2014,Nierenberg2017,Nierenberg2020}. In Section \ref{sec:methods} we explain our method of sampling the posterior distribution of our PBH parameters of interest using forward modeling. In Section \ref{sec:results}, we present the results of our modeling and comparison to real data, and in Section \ref{sec:theend} we discuss further expansions on this study.  When necessary, we use the cosmology parameters of \citet{Planck2020} throughout this analysis, although we stress that our results do not depend sensitively on this assumption.
	
\section{Methods}\label{sec:methods}
	
In this section, we first describe the goal of this paper, to obtain a posterior distribution on the PBH parameters of interest, which we achieve using an Approximate Bayesian Computing forward modeling method. We first model the lens substructure using the method developed in \cite{Gilman2019}, and then we model the effect of a possible PBH population.

\subsection{Inference}\label{subsec:inf}

We are striving to measure the posterior probability of dark matter model parameters; here our likelihood function $\mathcal{L}$ can be written as:

\begin{equation}	
    \mathcal{L}(D_i|\theta_{f,M}) = \int p(D_i|m_r, \theta_r)p(\theta_r,m_r|\theta_{f,M}) d m_r d \theta_r, 
    \label{likelihood}
\end{equation}
where $D_i$ is the observed image positions and flux ratios for a certain lens, $\theta_{f,M}$ represents our target model parameters, $M_{\mathrm{PBH}}$ and $f_{\mathrm{PBH}}$, $m_r$ is a certain lens model realization, and $\theta_r$ is the set of non-PBH model parameters that we marginalize over. We use the method described by \cite{Gilman2020} to sidestep evaluating this integral directly, which would require a computationally intractable exploration of a vast parameter space. This sidestepping is accomplished by forward modeling data, generating flux ratios from many sets of model parameters, and then comparing the results to the observed data via a summary statistic; from this process we can extract $\theta_{f,M}$ that represent our posterior probability distribution. This is an Approximate Bayesian Computing (ABC) method (\citealt{abc84}, see also, e.g., \citealt{abchandbook}) of creating a large set of stochastically varying simulated data and accepting simulations close to the real data to sample a posterior. ABC has been used in astrophysical forward-modeling problems where a direct calculation of the likelihood function is infeasible and data can be simulated; see, e.g., \cite{WeyantABC2013}, \cite{AkeretABC2015}, \cite{BirrerABC2017}. 

We use the sample of eleven quadruply-imaged quasars selected for flux ratio analysis by \citet[][Section 2.2]{Gilman2022} because the size of the source, either observed as O[III] emission from the narrow line region or CO (10-11) radio emission, is larger than the scale that would be affected by microlensing or image arrival time delay, and the main lensing galaxy does not require modeling for a known stellar disk component (\citealt{Hsueh2016}, \citealt{Hsueh2017}, \citealt{Gilman2017}). Photometry data used for each lens is referenced in Table~\ref{tab:params}. 

We generate a lens model using lenstronomy\footnote{\url{https://github.com/sibirrer/lenstronomy}} \citep{lenstronomy1, lenstronomy2}. The lens model is optimized to match the observed image position, with the added astrometric uncertainty. Any draw of parameters that does not match the observed image positions would be rejected in the posterior, so we reduce computation time by requiring the lens model fit the positions.

We compute the magnification, and thus the flux, of each image in our lens system model realization, then obtain the three flux ratios $r_{\mathrm{model}}$ between the four images. Only flux ratios are used because the intrinsic source brightness is not known. We compare the forward-modeled flux ratios to the observed flux ratios $r_{\mathrm{obs}}$ with the summary statistic 
\begin{equation}	
    S(r_{\mathrm{model}},r_{\mathrm{obs}}) = \sqrt{\sum_{\mathrm{i}=1}^{3}({r_{\mathrm{model}(\mathrm{i})}-r_{\mathrm{obs}(\mathrm{i})}})^2}. 
    \label{stat}
\end{equation}
We generate $\mathcal{O}10^5$ - $10^6$ lens model realizations sampling from our parameter space from which we choose the 1,500 lowest summary statistics to represent a sample of the posterior distribution. We construct a continous approximation of the likelihood function for each lens by applying a kernel density estimate to the accepted samples, and multiply the resulting likelihoods to obtain the final posterior.

\subsection{Model parameters}\label{subsec:params}

The lens and halo substructure modeling process follows from \cite{Gilman2020b}, \cite{Gilman2020}, and \cite{Gilman2019}. The lensing galaxy, or main deflector, is modeled as a power-law ellipsoid with external shear. The properties of the main deflector that are optimized during initial lens model fitting are the Einstein radius, centroid, ellipticity, ellipticity angle, and shear angle. If the main deflector has any known satellite galaxies, they are included in the model as a singular isothermal sphere mass profile. The main deflector mass $M_{\mathrm{host}}$, log profile slope $\gamma_{\mathrm{macro}}$, and shear $\gamma_{\mathrm{ext}}$ are sampled in the forward model.

Subhaloes are rendered from $10^6$-$10^{10}$M$_\odot$, from the lowest mass we are sensitive to to the highest mass of halo we expect to be entirely DM.  The projected mass density $\sum _{\mathrm{sub}}$ and power-law slope $\alpha$ parameterize the subhalo mass function (SHMF),
\begin{equation}	
    \dfrac{d^{2}N_{\mathrm{sub}}}{dmdA} = \dfrac{\Sigma_{\mathrm{sub}}}{m_{0}}\bigg(\dfrac{m}{m_{0}}\bigg)^{\alpha}\mathcal{F}(M_{\mathrm{halo}},z),
\end{equation}
    where $\mathcal{F}(M_{\mathrm{halo}},z)$ is a function to scale the number density of subhaloes with main lensing halo mass and redshift as described in \cite{Gilman2020}. The pivot mass $m_0$ is set to 10$^8$M$_\odot$ \citep{Fiacconi2016}. For the line-of-sight haloes, we use the Sheth-Torman halo mass function \citep{shethtorman} with two-halo term $\xi_{\mathrm{2halo}}$ as a scaling factor to account for correlated structure near the host halo (see \citealt{Gilman2019}) and $\delta_{\mathrm{los}}$ as an overall amplitude scaling factor:
\begin{equation}	
    \dfrac{d^{2}N_{\mathrm{los}}}{dmdV} = \delta_{\mathrm{los}}(1+\xi_{\mathrm{2halo}}(M_{\mathrm{halo}},z))\dfrac{d^{2}N}{dmdV}|_{\mathrm{ShethTormen}}.
\end{equation}
Given the PBH mass is distributed along with CDM subhaloes, the Sheth-Torman mass function should be broadly applicable, but there may be an enhancement of the power spectrum on small scales caused by isocurvature perturbations from PBH \citep{Afshordi2003, Gong2017}. Our constraint is more conservative because we do not take this enhancement into account. The free model parameters and priors are as follows:
\begin{itemize}
    \item $M_{\mathrm{PBH}}$ [M$_\odot$], the PBH monochromatic mass, with a prior of $10^4$-$10^6$M$_\odot$ chosen to include PBH that are large enough to affect the flux ratios given the background source size but not larger than the minimum rendered halo mass;
    \item $f_{\mathrm{PBH}}$, the  PBH mass fraction of total DM, with a prior of 0-50\%;
    \item $\Sigma _{\mathrm{sub}}$, the SHMF normalization, with a prior of 0-0.1 kpc$^{-2}$. We allow broad uncertainty in the number of subhaloes to account for uncertainties associated with tidal stripping; 
    \item $\alpha$, the log slope of SHMF, with a prior ranging from -1.85 to -1.95 as predicted by $\Lambda$CDM N-body simulations (\citealt{Springel2008}, \citealt{Fiacconi2016});
    \item $\delta_{\mathrm{los}}$, the line-of-sight halo mass function scaling factor, with a prior of 0.8-1.2 that accounts for differences between theoretical models of the halo mass function \citep[e.g.][]{Despali2016} and uncertainties in cosmological parameters;
    \item $\gamma_{\mathrm{macro}}$, the log slope of main deflector mass profile, with a data-motivated prior of 1.9-2.2 (\citealt{Auger2010});
    \item $\sigma _{\mathrm{source}}$, the background source size, differing depending on whether the source is observed in narrow-line \citep{Muller-Sanchez2011} or other regions \citep{Chiba2005,Stacey2020} surrounding the background quasar listed for each lens in Table \ref{tab:params} ;
    \item $M_{\mathrm{host}}$, the mass of the main lens host galaxy (see Table \ref{tab:params} and \cite{Gilman2020} for a discussion of these priors constructed from individual lens data);
    \item $\gamma_{\mathrm{ext}}$, the external shear in the main lens plane (see Table \ref{tab:params}, with ranges based on the individual lens data determined in \cite{Gilman2022});
    \item $\delta_{\mathrm{xy}}$ [milliarcsec], the image position uncertainty;
    \item $\delta_{\mathrm{f}}$, the image flux uncertainty.
\end{itemize}
References to photometric measurement information are listed in Table~\ref{tab:params}. Our target parameters are $M_{\mathrm{PBH}}$ and $f_{\mathrm{PBH}}$, and we marginalize over the others when they are sampled together in the posterior. For each realization, the model parameters are drawn from a prior distribution and the halo placement is stochastic. Some lenses have photometrically-estimated redshifts \citep{Gilman2020} so we sample the redshift probability distribution function and marginalize over it for those lenses.

Lens RXJ1131+1231 was modeled with two Gaussian source components to match the data of \cite{Sugai2007}. Lenses with an imaged satellite companion are modeled with the companion in the source plane as a single isothermal sphere with position uncertainty of 50 milliarcsec.

\begin{table*}
 \centering
    \caption{Priors for the parameters in our model that are lens-dependent. For a description of all free parameters, including those with priors shared between all lenses, see text. The host halo mass $M_{\mathrm{host}}$ has a Gaussian prior and other priors are uniformly distributed. The rightmost column has the reference for the photometry data.}
    \label{tab:params}
    \begin{tabular}{ccccc}
    \hline
    {Lens Name} & {$\sigma _{source}$ [pc]} & {$M_{host}$ [M$_\odot$] ($\mu$, $\sigma$)} & {$\gamma_{ext}$} &  {Ref.}\\
    \hline
    B1422+231 & 25-60 & 13.3, 0.3  & 0.08-0.4    & $^a$ \\ 	 %
    HE0435-1223 & 25-60  & 13.2, 0.3  & 0.015-0.15     & $b$  \\ 	 %
    MG0414+0513 & 5-15  & 13.5, 0.3  & 0.01-0.32     & $c$ $d$  \\ 	 %
    PG 1115+080 & 5-10  & 13.0, 0.3  & 0.002-0.12   & $e$  \\ 	 %
    PS J1606-2333 & 25-60  & 13.3, 0.3  & 0.1-0.28 & $f$  \\ 	 %
    RX J0911+0551 & 25-60  & 13.1, 0.3  & 0.05-0.25    & $f$  \\ 	 %
    RX J1131-1231 & 25-80  & 13.9, 0.3  & 0.06-0.28   & $g$  \\ 	 %
    WFI 2026-4536 & 25-60  & 13.3, 0.3  &  0.015-0.16   & $f$  \\ 	 %
    WFI 2033-4723 & 25-60  & 13.3, 0.3  & 0.07-0.26   & $f$  \\ 	 %
    WGD J0405-3308 & 25-60  & 13.3, 0.3  & 0.0025-0.12   & $f$  \\ 	 %		   
    WGD 2038-4008 & 25-60  & 13.04, 0.15$^*$ & 0.005-0.08  & $f$  \\ 	 %  
    \hline
    \end{tabular} \\
    $^*$This value from \cite{Anowar2022} \\
    \textbf{References} ---
    $^a$\cite{Nierenberg2014}
        $^b${\cite{Nierenberg2017}}
        $^c${\cite{Stacey2018}}
        $^d${\cite{Stacey2020}}
        $^e${\cite{Chiba2005}} \\
        $^f${\cite{Nierenberg2020}}
        $^g${\cite{Sugai2007}}
\end{table*}

\subsection{PBH deflection modeling}\label{subsec:pbh}

If some fraction $f_{\mathrm{PBH}}$ of dark matter exists in the form primordial black holes, the distribution of these objects should follow a population of haloes with Navarro-Frenk-White (NFW) profiles \citep{NFW}. Thus, the first step in our analysis is to generate a population of NFW haloes and subhaloes throughout the lensing volume. We create a realization of DM haloes and subhaloes using the lenstronomy affiliate package pyHalo\footnote{\url{https://github.com/dangilman/pyHalo}} \citep{pyHalo}.  We can calculate the mass fraction of dark matter rendered in haloes $f_{\mathrm{halo}}$, which we use to determine the number of PBH that is clustered with the halo mass, and a stochastic distribution of line-of-sight and lens-plane subhalo masses, which we use to determine the clustered PBH position. 

\begin{figure}
    \includegraphics[width=\columnwidth]{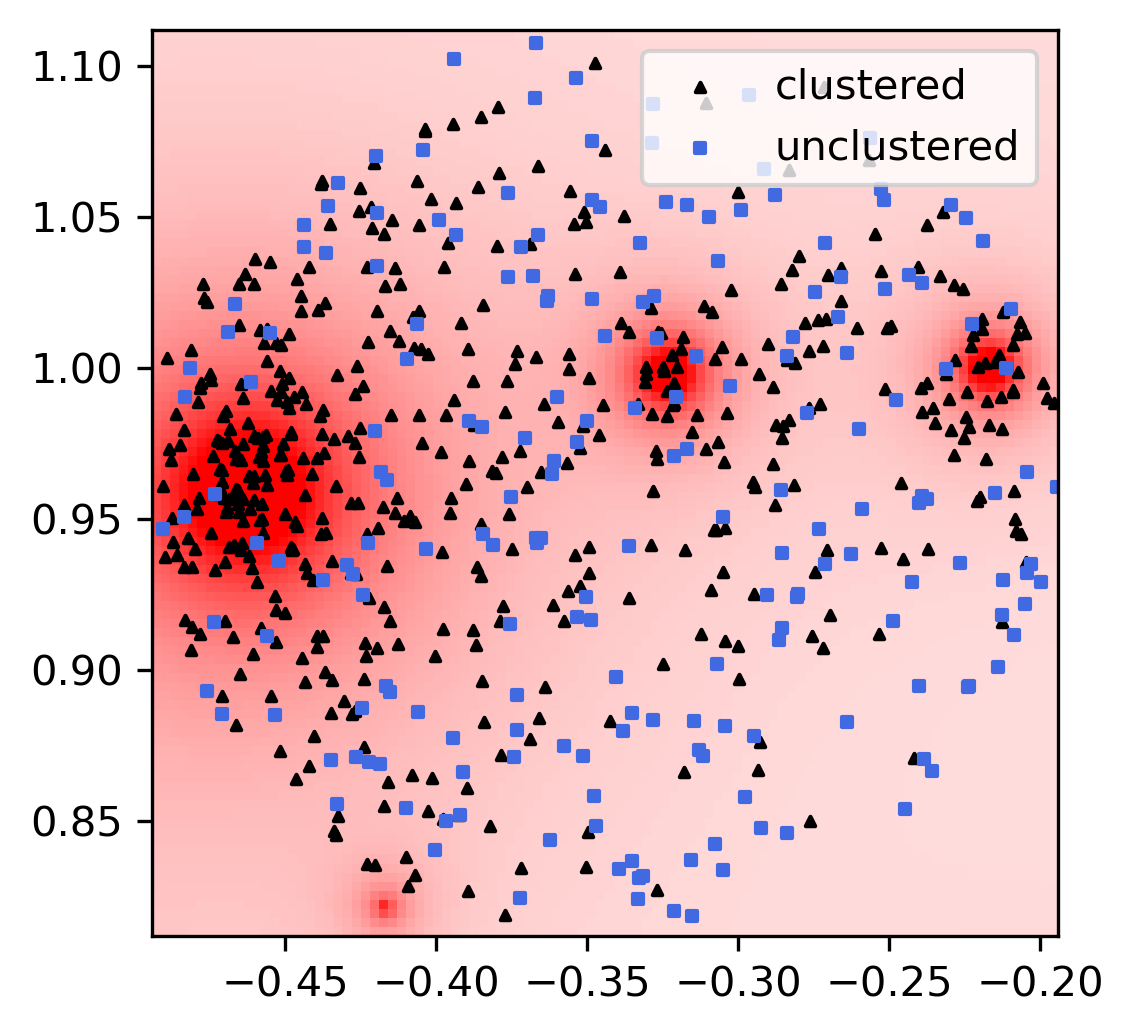}
    \caption{An example of PBH positions in a single lens plane. In this figure, $M_{\mathrm{PBH}}$ is 10$^4$ M$_\odot$ and $f_{\mathrm{PBH}}$ is 0.5. The axes are in arcseconds and the background colormap intensity varies linearly with the projected mass in dark matter. The unclustered population (blue squares) is distributed uniformly across the rendering aperture, while the clustered population (black triangles) tracks the projected dark matter mass density in haloes. \label{PBHmap}}
\end{figure}

\begin{figure*}
    \includegraphics[width=\columnwidth]{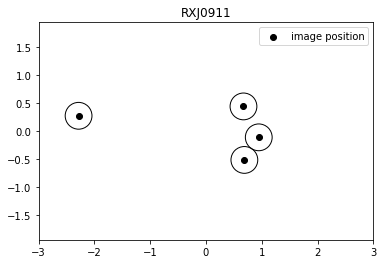}
        \includegraphics[width=\columnwidth]{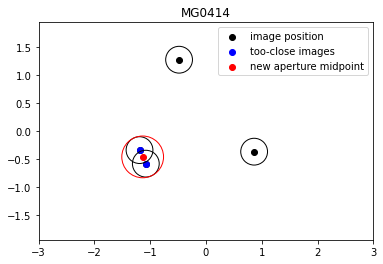}
    \caption{Visualization of rendering area relative to image position. The circles around the image positions represent the rendering area for lensing substructure. On the left, for lens RX J0911+0551, these areas do not overlap for the chosen radius of 0.24". On the right, there is significant overlap between the rendering areas for two images, so a new aperture is drawn around both images to avoid double-placement of PBH in the overlap region.   \label{circles}}
\end{figure*}

To determine the spatial distributiion of the black holes at each redshift plane along the line of sight, we first compute the projected mass in dark matter at the lens plane from the population of NFW haloes distributed throughout the lensing volume. If we were to render haloes down to the minimum halo mass in CDM, then all of the PBH would track the density of dark matter in haloes. However, as we only render a fraction of the total mass of dark matter in haloes, a number $N_{\mathrm{clustered}} = (f_{\mathrm{PBH}})(f_{\mathrm{halo}})(\rho_{\mathrm{DM}}(z))(V/M_{\mathrm{PBH}})$ will track the dark matter density in haloes. We distribute this population of PBH with a spatial probability density that varies in proportion with the project mass in dark matter at each lens plane, as illustrated by Fig.~ \ref{PBHmap}. The mass added in PBH is removed from the rendered particle-DM subhalo mass. We place the remaining $N_{\mathrm{smooth}}$ point masses randomly across each redshift plane, tracking the smooth background distribution of dark matter that we do not place in haloes. For each image of the lens, we add PBH at discrete redshift steps along the line of sight within a circular aperture of 0.24" for 10$^8$ M$_\odot$, scaling with the root $M_{\mathrm{PBH}}$ to a minimum of 0.15". For images closer than 0.24", we add half the distance between the two points to the aperture and centre it at the midpoint between the images, treating the rendering area for those images as one aperture as illustrated in Fig.~ \ref{circles}.

\begin{figure*}
    \includegraphics[trim=0.25cm 1cm 0.25cm
	1cm,width=\columnwidth]{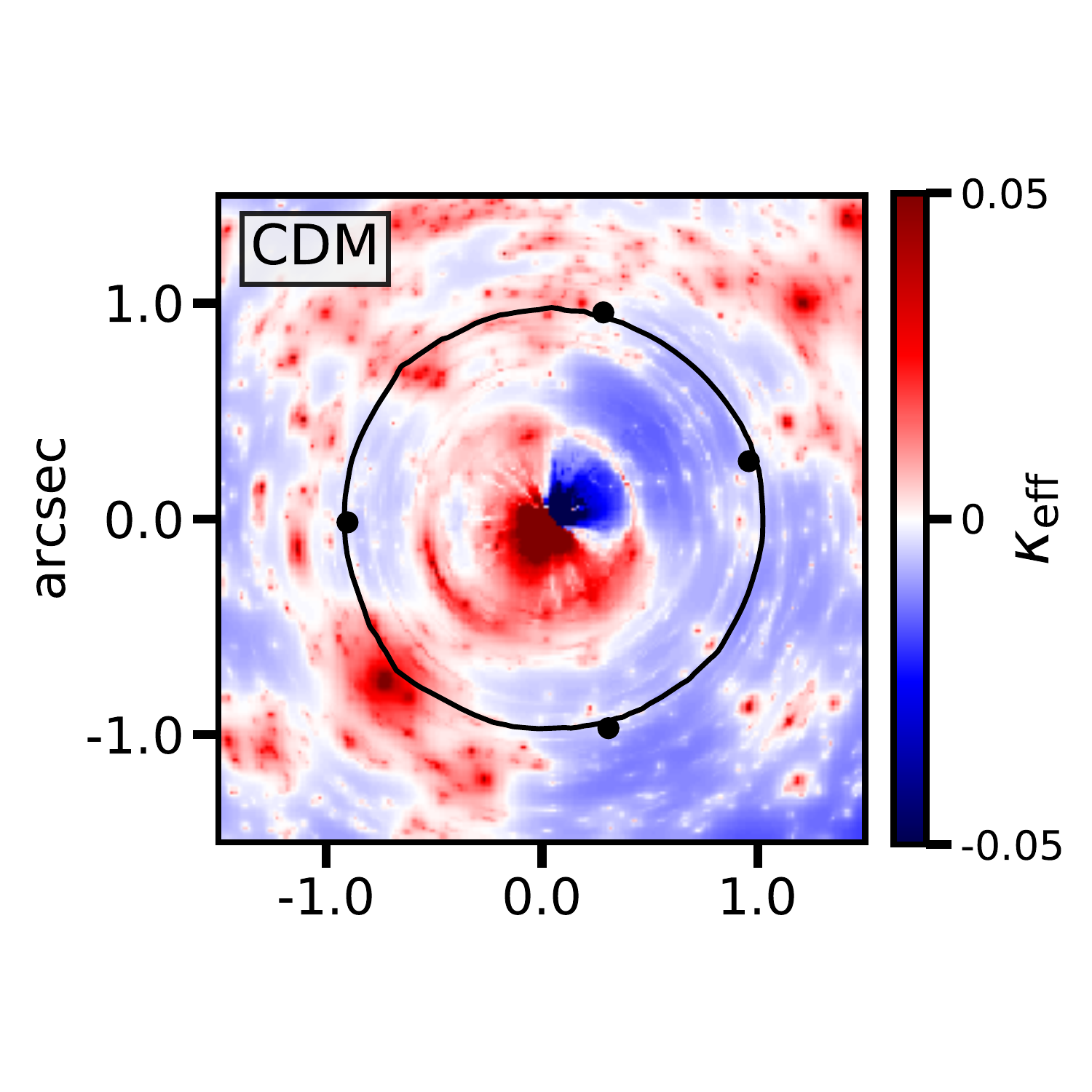}
    \includegraphics[trim=0.25cm 1cm 0.25cm
	1cm,width=\columnwidth]{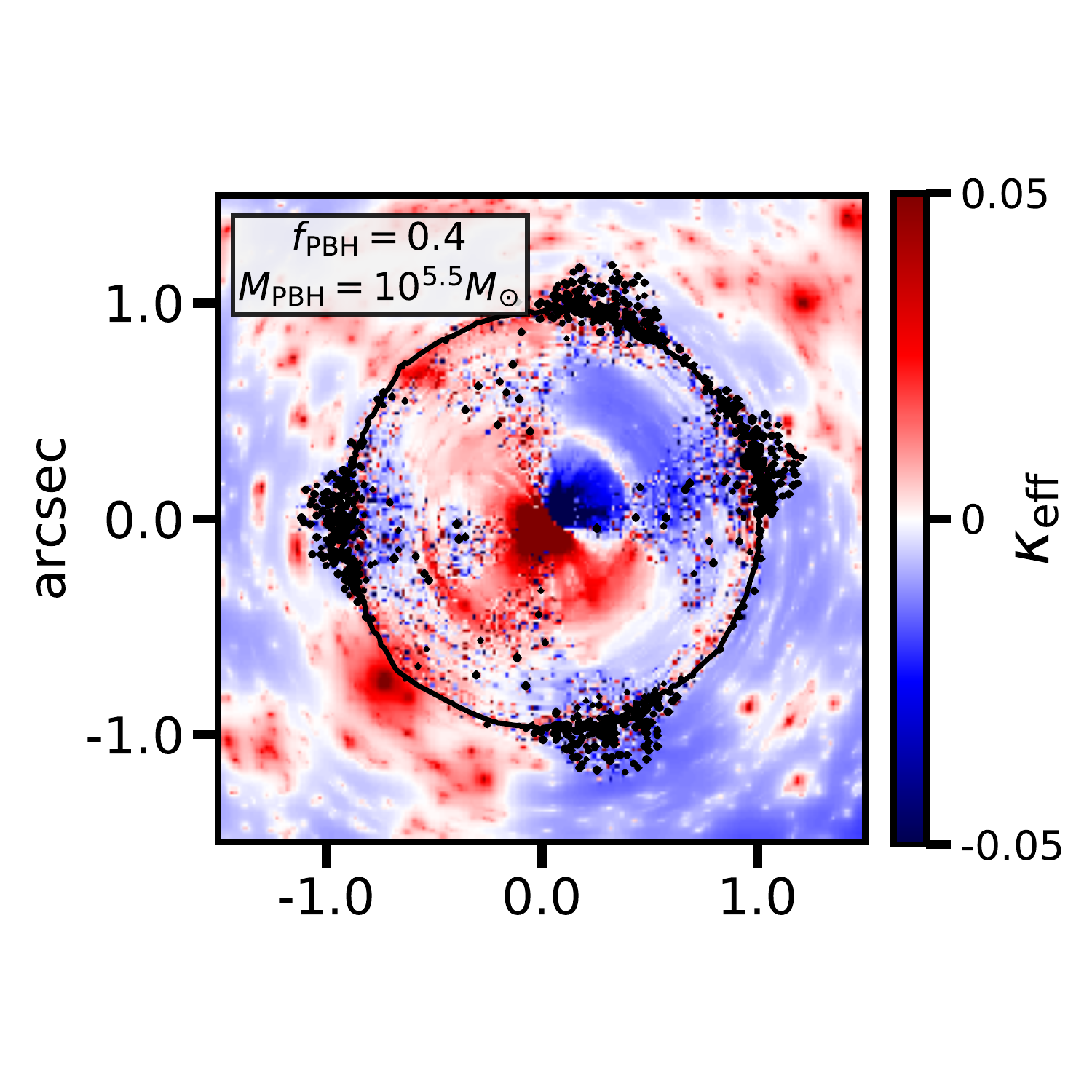}
    \caption{Effective multiplane convergence, a two-dimensional representation of a full population of line of sight haloes and subhaloes, for a dark matter realization in CDM (left) and with PBH substructure (right). Red corresponds to a density higher to that of the mean dark matter density, while blue corresponds to an underdensity. Black circles are plotted at each of the four quad image positions, and the black curves are the critical curves, which follows the region of maximum image magnification. Small-scale features in the convergence map that appear to track towards the origin are associated with black holes rendered around the path followed by the lensed light rays. Deformation of the critical curve by the PBH population suggests they will strongly perturb image flux ratios. \label{empc}}
\end{figure*}

In Fig.~ \ref{empc} we have plotted effective multiplane convergence for a lens model with $M_{\mathrm{PBH}} = 10^{5.5}$M$_\odot$ and $f_{\mathrm{PBH}} = 0.4$ alongside that of a model with no PBH substructure. This effective multplane convergence is defined as the multiplane convergence (half the divergence of the effective deflection through the lensing planes $\boldsymbol{\alpha}_{\mathrm{eff}}$) from the lens model minus the macromodel convergence, thus $\kappa_{\mathrm{eff}} \equiv \frac{1}{2} \nabla \cdot \boldsymbol{\alpha}_{\mathrm{eff}} - \kappa_{\mathrm{macro}}$. On a convergence map, which corresponds to surface density, the point masses produce markedly different lensing signatures to the less centrally concentrated NFW profiles. 

After the PBH have been distributed in the lens model, the deflection from the new point masses is accounted for by re-fitting the lens model to the observed image positions. We raytrace through this final lens model to get the simulated image flux ratios, and calculate our summary statistic for the realization.


\section{Results and comparison with previous work}\label{sec:results}

\begin{figure}
    \includegraphics[width=\columnwidth]{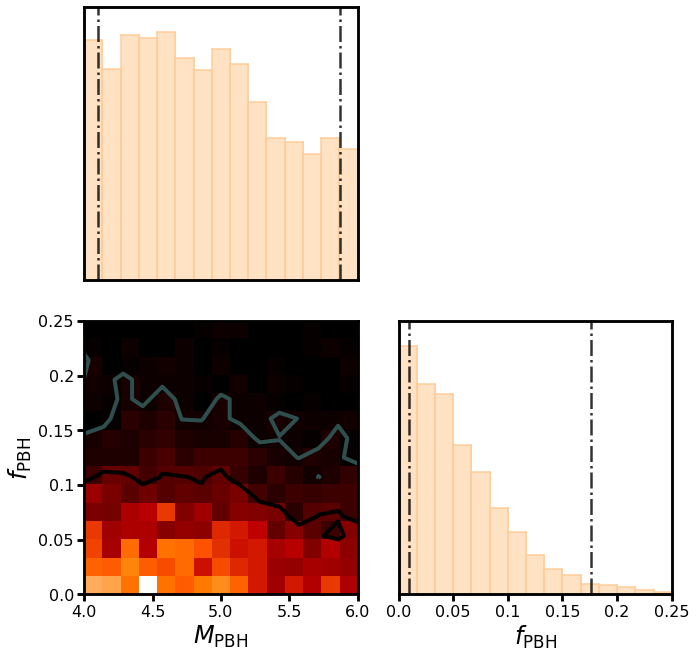}
    \caption{Joint posterior distribution of the PBH mass and mass fraction obtained from analyzing eleven strong lenses, marginalized over $\Sigma _{\mathrm{sub}}$, $\alpha$, and $\delta_{\mathrm{los}}$. The vertical dot-dash lines in the panels showning marginal likelihoods represent 95\% confidence intervals. The lighter contours are 95\% confidence region and the darker contours bound the 68\% confidence region. \label{combinedreal}}
\end{figure}

We present our constraint on PBH dark matter from the posterior distribution of our target PBH parameters for 11 lenses in (Fig.~ \ref{combinedreal}), which were combined and marginalized over the main deflector and subhalo parameters described in Section \ref{subsec:params}. We obtained a 95\% upper limit on $f_{\mathrm{PBH}}$ of 0.17 accross the probed mass range. We see there is a tentative anticorrelation between $M_{\mathrm{PBH}}$ and $f_{\mathrm{PBH}}$, as we would expect. 

\begin{figure}
   \includegraphics[width=\columnwidth]{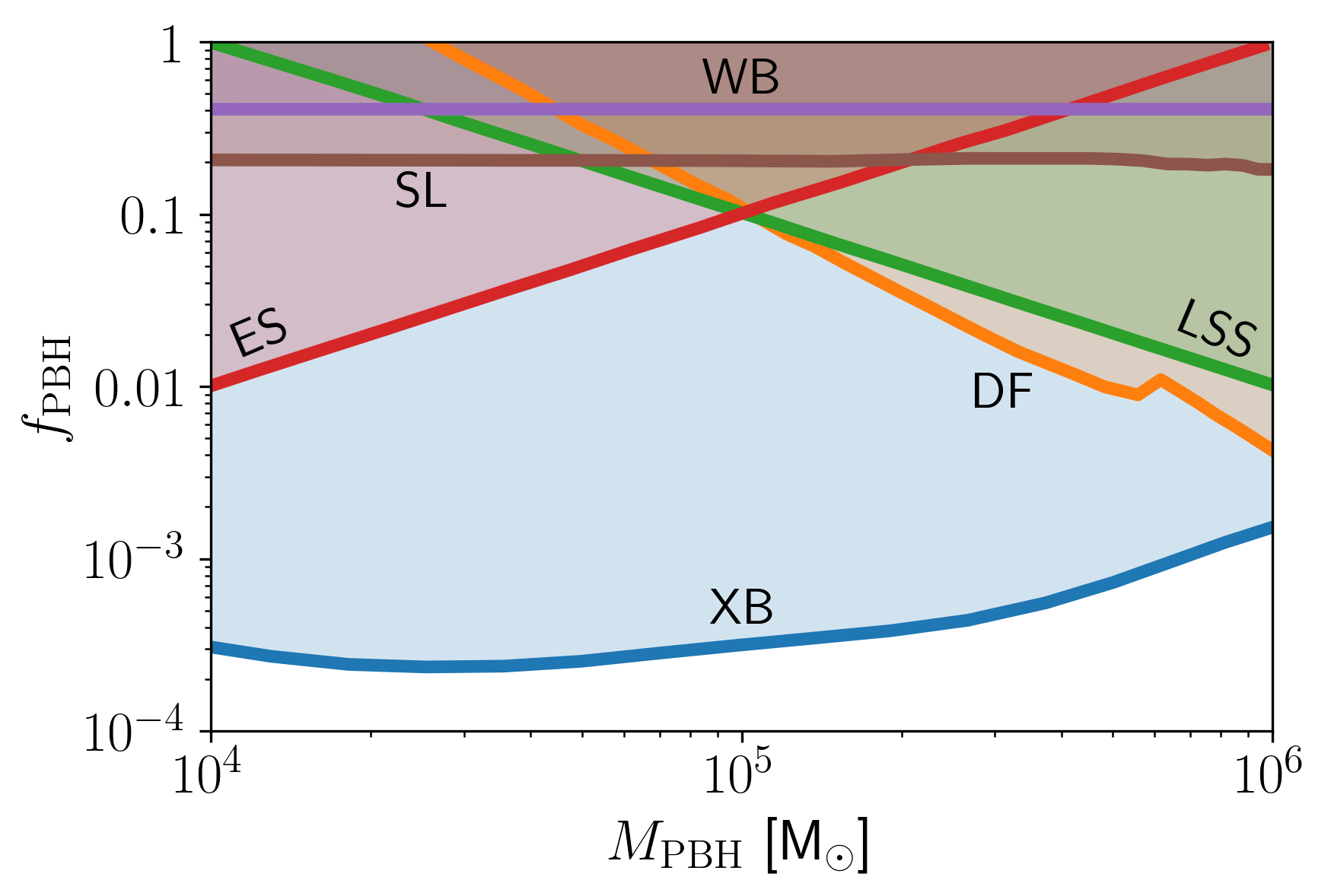}
    \caption{Constraints from disruption of wide binaries (WB) \citep{Quinn2009}, Eridanus II star cluster surviving possible destruction by dynamical heating (ES) \citep{Brandt2016}, halo dynamical friction (DF) \citep{CarrDF1999}, large-scale structure (LSS) (\citealt{Afshordi2003}, \citealt{Mack2007}), X-ray background from accretion (XB) \citep{InoueKusenko2017}, and our constraint from strong lensing flux ratio analysis (SL). \label{exclusion}}
\end{figure}

Our constraint is plotted along with others in the same mass range in Fig.~ \ref{exclusion}. The constraint is stronger than that placed by radial velocity measurements of three wide binary systems that could be disrupted by a PBH population \citep{Quinn2009}, but it is partially within the bounds of the other four constraints. However, our method is totally independent of the other bounds, and thus provides an important cross-check of the assumptions of other methods, and their potential systematic uncertainties. 

The X-ray accretion background constraint depends sensitively on assumption about the physics of gas accretion on to PBH and the possible subsequent formation of an accretion disk, the density of the interstellar medium (ISM), and PBH motion through the ISM. The constraint shown from \cite{Brandt2016}, similar to that of \cite{Quinn2009}, is from the survival of the Eridanus II star cluster that would be dynamically heated into dispersal by PBH dark matter. This assumes that the Eridanus II cluster formed in place. The dynamical constraint placed by \cite{CarrDF1999} assumes that PBH will drift to the centres of galaxies, but this has been argued to be avoidable if PBH are regularly dynamically ejected as well \citep{Xu1994}. The large-scale structure constraint in Figure~\ref{exclusion} is from the effect of PBH on the matter power spectrum as probed by the Lyman-$\alpha$ forest, which in turn depends on assumptions and modeling of its thermodynamics \citep{Villasenor22,Viel2013}. 

Finally, as shown by \cite{Banik2021}, N-body simulations with dark matter particle masses comparable to the PBH mass range we consider result in small-scale perturbations to stellar streams. Interpreting this result in the context of primordial black holes suggests streams also can constrain the contribution of PBH to the dark matter. 

\section{Discussion and conclusions}\label{sec:theend}

We develop a new method for including PBH substructure in a lens model for flux ratio analysis, and present independent constraints on the fraction of dark matter that could be composed of relatively massive primordial black holes. We obtained a constraint on $f_{\mathrm{PBH}}$ less than ~0.17 for $M_{\mathrm{PBH}} = 10^4$-$10^6$M$_\odot$ (95\% C.L.). The mass distribution for the PBH in this work is monochromatic as a conservative constraint, but the limit can be converted to an arbitrary extended mass distribution via the method presented in \cite{Carr2017}. This constraint is totally independent of others in the same mass range. 

In the spirit of a first application of this method, we make several simplifying assumptions throughout this process. We do not account for the effect of PBH formation on the assembly history of subhaloes and how that could possibly affect the mass functions and density profiles that we are also assuming. We allow for a very general parameterized form of these functions and marginalize over the parameter space to reduce the rigidity of our models.  As samples of quads improve and our method becomes more constraining, we will revisit the simplifying assumptions.

In the future, these constraints will be improved by applying the method to larger samples of lenses that are currently being discovered \citep{Schmidt22} and will be discovered in wide field surveys such as the Vera C. Rubin, Euclid, and Roman Observatories \citep[e.g.][]{Oguri2010}. Lens systems can also be followed-up with adaptive optics assisted instruments from the ground \citep{LIGER,KAPA}. Forthcoming data from the \textit{James Webb Space Telescope (JWST)} \citep{NierenbergJWST} will allow us to push to lower PBH mass scales because \textit{JWST} will measure flux ratios in the mid-infrared. This emission comes from a more spatially compact ($\sim 1 - 10 \ \rm{pc}$) region around the background source. The minimum deflection angle that impacts our data is determined by the size of the source, so the more compact source size will allow us to push to lower PBH mass scales than we can currently measure. 

\section*{Acknowledgements}

The authors thank Alex Kusenko for insightful and stimulating conversations, and Xiaolong Du for a productive discussion on halo mass functions.

This material is based upon work supported by the National Science Foundation Graduate Research Fellowship Program under Grant No. DGE-1650604. Any opinions, findings, and conclusions or recommendations expressed in this material are those of the authors and do not necessarily reflect the views of the National Science Foundation.
	
V.D. thanks the LSSTC Data Science Fellowship Program, which is funded by LSSTC, NSF Cybertraining Grant \#1829740, the Brinson Foundation, and the Moore Foundation; her participation in the program has benefited this work.

D.G. was partially supported by a HQP grant from the McDonald Institute (reference number HQP 2019-4-2).

T.T. acknowledges support by NSF through grants NSF-AST-1714953, NST-AST-1836016, NSF-AST-2205100 and by the Gordon and Betty Moore Foundation through grant 8548.

This work used computational and storage services associated with the Hoffman2 Shared Cluster provided by UCLA Institute for Digital Research and Education’s Research Technology Group.

This research is based on measurements made with the NASA/ESA Hubble Space Telescope obtained from the Space Telescope Science Institute, which is operated by the Association of Universities for Research in Astronomy, Inc., under NASA contract NAS 5–26555. These observations are associated with programs GO-15177 and GO-13732. Some of the measurements used herein were obtained at the W. M. Keck Observatory, which is operated as a scientific partnership among the California Institute of Technology, the University of California and the National Aeronautics and Space Administration. The Observatory was made possible by the generous financial support of the W. M. Keck Foundation.  The authors wish to recognize and acknowledge the very significant cultural role and reverence that the summit of Maunakea has always had within the indigenous Hawaiian community.  We are most fortunate to have the opportunity to conduct observations from this mountain.

%\facility{}
We acknowledge the use of the following software: Astropy (\citealt{astropy1} \& \citealt{astropy2}), \textsc{Colossus} \citep{colossus}, lenstronomy \citep{lenstronomy1, lenstronomy2} Matplotlib \citep{matplotlib}, NumPy \citep{numpy}, pandas \citep{pandas}, pyHalo \citep{pyHalo}, SciPy \citep{scipy}.

%%%%%%%%%%%%%%%%%%%%%%%%%%%%%%%%%%%%%%%%%%%%%%%%%%
\section*{Data Availability}
The data underlying this article will be shared on reasonable request to the corresponding author.


\bibliographystyle{mnras}
\bibliography{PBHpaper}

\appendix

\section{Testing the Pipeline}\label{sec:test}

\begin{figure*}
    \includegraphics[width=\columnwidth]{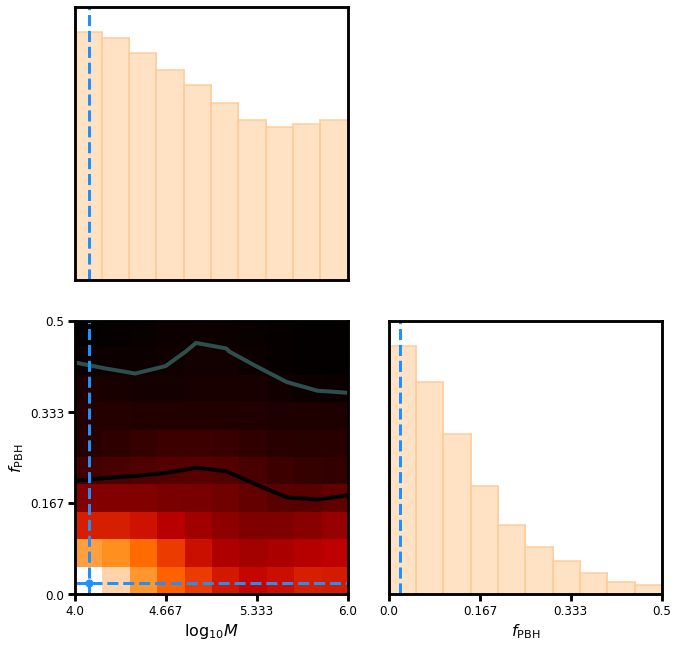}
        \includegraphics[width=\columnwidth]{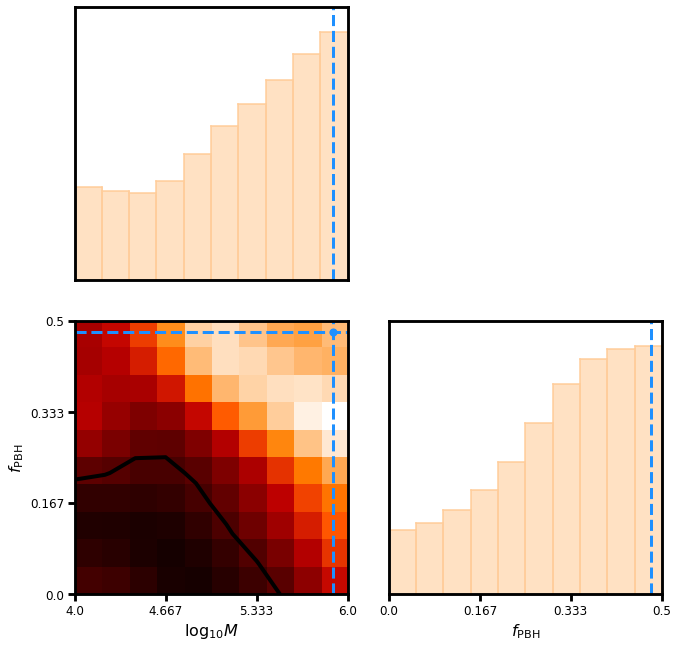}
    \caption{Posterior distributions created from simulated data using image positions and lensing priors of B1422+231. The posteriors are drawn from the 250 closest samples to the simulated "truth" flux ratios represented by dashed blue lines and corresponding to $M_{\mathrm{PBH}} = 10^{4.1}$M$_\odot$, $f_{\mathrm{PBH}} = 0.02$ on the left and  $M_{\mathrm{PBH}} = 10^{5.9}$M$_\odot$, $f_{\mathrm{PBH}} = 0.48$ on the right. \label{b1422triplot}}
\end{figure*}

\begin{figure*}
   \includegraphics[width=\columnwidth]{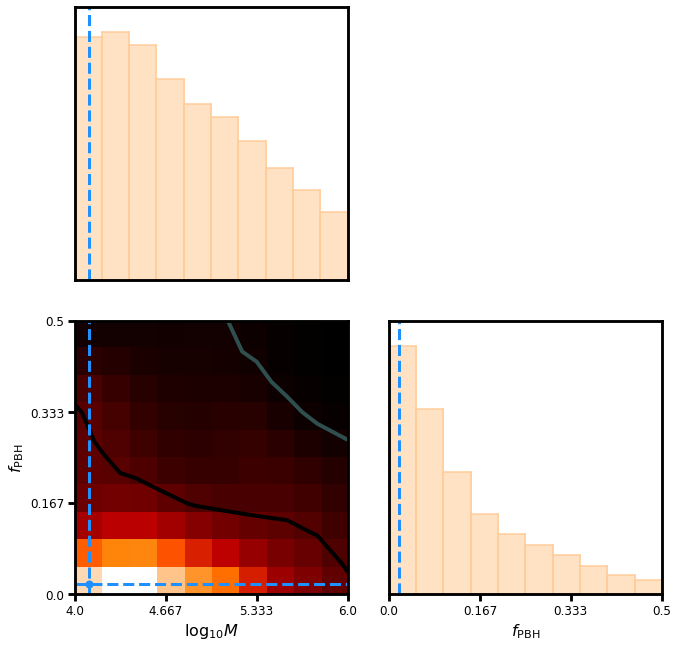}
        \includegraphics[width=\columnwidth]{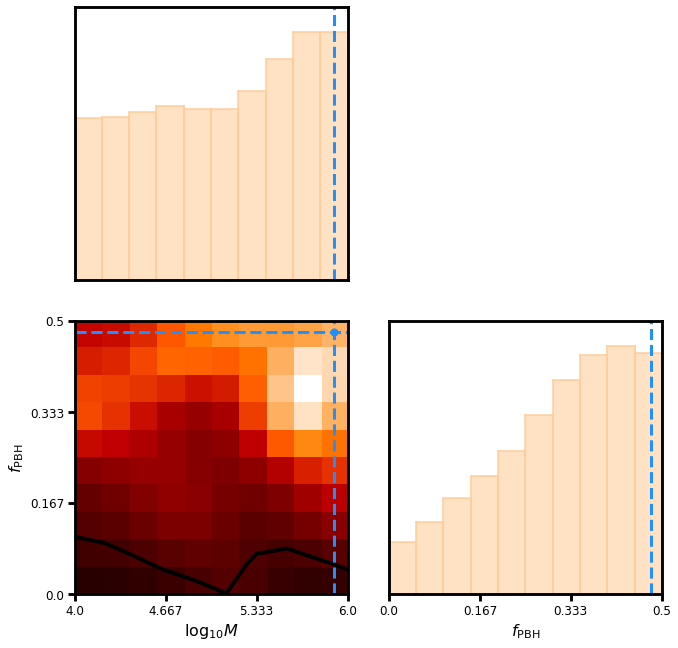}
    \caption{Product of posterior distributions based on image positions and priors of the three lenses B1422+231, PS J1606-2333, and WGD J2038-4008. As in Fig.~ \ref{b1422triplot}, the selected "true" flux ratios used to obtain each distribution are $M_{\mathrm{PBH}} = 10^{4.1}$M$_\odot$, $f_{\mathrm{PBH}} = 0.02$ on the left and  $M_{\mathrm{PBH}} = 10^{5.9}$M$_\odot$, $f_{\mathrm{PBH}} = 0.48$ on the right. \label{combinedsim}}
\end{figure*}

Using 50,000 simulated lens model realizations of B1422+231, we tested the performance of our method by applying it to simulated data. We choose a realization with a low target mass and mass fraction of PBH and used the simulated flux ratios as the "true" flux ratios in the computation of the summary statistic. From this, we obtain the posterior distributions shown in Fig.~ \ref{b1422triplot}. We repeat the exercise using a high PBH mass and mass fraction, and show the resulting inference on the right of Fig.~ \ref{b1422triplot}. The other parameters described in Section \ref{sec:methods} were fixed in the middle of their uniform prior ranges. This process was carried out similarly for the lenses PS J1606-2333 and WGD J2038-4008, and the marginalized joint posterior distribution of all three lenses is shown in Fig.~ \ref{combinedsim}.

% Don't change these lines
\bsp	% typesetting comment
\label{lastpage}
\end{document}